\documentclass[aps,prl,twocolumn,showpacs]{revtex4}
\usepackage{amssymb}
\usepackage{amsmath}
\usepackage{epsfig}
\usepackage{color}

\begin{document}

\title{Parametric resonance induced chaos in magnetic damped driven pendulum}
\author {Giorgi Khomeriki}
\affiliation {Vekua school of Physics \& Mathematics, 9 Tchaikovsky,
0105 Tbilisi, Georgia}

\begin{abstract}
\noindent A damped driven pendulum  with a magnetic driving force,
appearing from a solenoid, where ac current flows is considered. The
solenoid acts on the magnet, which is located at a free end of the
pendulum. In this system the existence and interrelation of chaos
and parametric resonance is theoretically examined. Derived
analytical results are supported by numerical simulations and
conducted experiments.
\end{abstract}

\pacs{05.45.Ac, 42.65.Yj, 05.45.-a}

\maketitle

\noindent{\it Introduction:} Chaos in damped driven pendulum system
has a long standing history (see e.g. Refs. \cite{Gregory1,Gregory2}
and references therein) and is applicable in vast variety of
condensed matter problems \cite{Malomed,Ruffo,Alekseev1} ranging
from Josephson junctions to easy-plane ferromagnets. Governing
equation is written in the standard form:
\begin{eqnarray}
\ddot{\alpha}=-{\Omega^2}\sin\alpha-q\dot{\alpha}+{f_D}\sin{\omega t}
\label{1}
\end{eqnarray}
where $\Omega$ and $q$ coefficients are usually fixed and $f_D$ is a
one we control. Increasing control parameter $f_D$ period doubling
\cite{Yorke,Bevivino} bifurcation scenario and transition to chaos
takes place \cite{Hubbard,Alekseev2,Harish}. In all the mentioned
papers control parameter is constant \cite{Hauptfleisch} or a
driving force has a time periodic singular character (kicked excited
systems \cite{Popov}). In the present paper driving force is
position angle $\alpha$ dependent, particularly, here, a realistic
example of driven damped pendulum model is considered. In this context, driving force is of a magnetic origin, particularly a
solenoid with ac current is acting on the magnet, which plays a role
of a bob in a pendulum with a rigid rod (see Fig. 1). Therefore the
amplitude of a harmonic force $f_D$ greatly depends on the distance
between solenoid and the magnet, making it angle dependent in a
non-trivial manner.

In the frames of the model \eqref{1} a possibility of onset of chaos
has been examined analytically, numerically and experimentally. The
similar model of magnetic pendulum has been studied long before
\cite{doubochinski}, particularly, different orientation of solenoid
and magnet has been considered, where the orientation of the
solenoid is perpendicular to the pendulum's rod when the deviation
angle is zero. In this case one gets quantization of amplitudes with
no indication of onset of chaos, while in our case with parallel
orientations of solenoid and pendulum in unperturbed position (see
again Fig. 1) for some values of ac field and/or distance between
solenoid and magnet chaos is observed due to the parametric
resonance \cite{Belyakov}. Thus the main peculiarity of our model is
that the existence of parametric resonance is a necessary condition
for the onset of chaos in the system.

\begin{figure}[t]
\includegraphics[scale=.6]{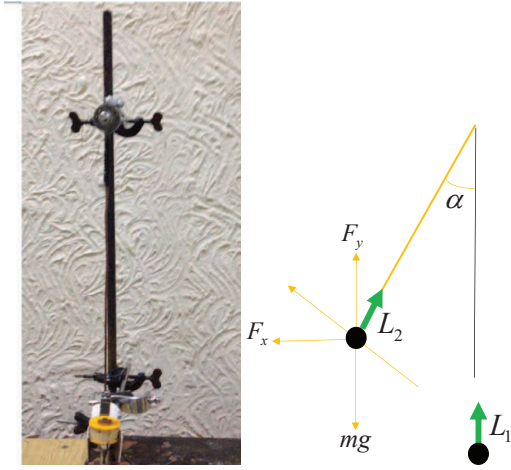}
\caption{Experimental setup (left) and schematics (right) for driven
damped magnetic pendulum. $L_1$ is a dipolar moment of solenoid,
$L_2$ is a dipolar moment of the magnet. $\alpha$ is a deviation
angle of the pendulum from vertical. $F_x$ and $F_y$ are $x$ and $y$
components of magnetic force acting on the magnet. } \label{fig_01}
\end{figure}

\noindent{\it Theoretical Model:} In my experiments and numerical
simulations the magnet is rigidly fixed in the place of a bob of the
pendulum in such a way that the directions of its magnetic moment
and the rod of pendulum coincides. Approximating solenoid and magnet
as point-like magnetic moments ($\overrightarrow{L}_1$ and
$\overrightarrow{L}_2$, respectively), one can readily write down
their dipolar interaction energy as follows:
\begin{eqnarray} \nonumber
U=\frac{\mu_0}{4\pi}\left(\frac{3\cdot(\overrightarrow{L}_1\cdot\overrightarrow{r})(\overrightarrow{L}_2\cdot
\overrightarrow{r})}{{r^{5}}}-\frac{\overrightarrow{L}_1\cdot\overrightarrow{L}_2}{r^{3}}\right)
\label{200}
\end{eqnarray}
where $\overrightarrow{r}\equiv(x,~y)$ is a radius vector of magnet
with respect to the solenoid, $r=\sqrt{x^{2}+y^{2}}$. Taking into
account now that ac current is flowing into the solenoid and the
magnet is attached at the free end of the pendulum one can write for
the components of magnetic moments following expressions (see also
Fig. \ref{fig_01}):
\begin{eqnarray}
{L_{1x}}=0, \quad {L_{2x}}=-{L_2}\frac{x}{\ell},
\nonumber \\
{L_{1y}}={L_1(t)}, \quad   {L_{2y}}={L_2}\frac{{r_0}+\ell-y}{\ell} \label{2}
\end{eqnarray}
where $\ell$ is the length of the pendulum and $r_0$ is distance
between magnet and solenoid when the deviation angle from vertical
direction is zero (that is a minimal distance position between
solenoid and magnet).

Plugging \eqref{2} into \eqref{1} we find $F_x$ and
$F_y$ components of the forces acting on the magnet:
\begin{eqnarray}{F_x}=-\frac{\partial{U}}{\partial{x}}
\qquad {F_y}=-\frac{\partial{U}}{\partial{y}} \nonumber
\end{eqnarray}
we write Newton's second law for tangential axis of the pendulum as follows:
\begin{eqnarray}
m\ddot{\alpha}\ell={F_x}\cos{\alpha}+{F_y}\sin{\alpha}-mg\sin{\alpha}-q\dot{\alpha}
\label{3}
\end{eqnarray}
where a damping proportional to velocity has been included and $m$
is a mass of the magnet. We do not write here explicit expressions
for components of the force because of their cumbersomeness,
although their complete expressions will be used for numerical
simulations, while for analytics we just linearize \eqref{3} for
small deviation angles $\alpha$ and approximate $r\rightarrow r_0$:
\begin{eqnarray}
\ddot{\alpha}=-{\alpha}\left(\frac{g}{\ell}+\frac{12{L_1}(t){L_2}}{mr_0^{5}}+\frac{2{L_1}(t){L_2}}{m{\ell^{2}}{r_0^{3}}}\right)-q\dot{\alpha}
\label{4}
\end{eqnarray}
where ${L_1}(t)\equiv {L^0_1}\cos{2\omega t}$ because of the ac
current (with $2\omega$ frequency) flowing through the solenoid.
Then let us denote
\begin{eqnarray}
\omega_0=\sqrt{\frac{g}{\ell}}, \qquad
h={L^0_1}\left(\frac{12{L_2}}{mr_0^{5}}+\frac{2{L_2}}{m{\ell^{2}}{r_0^{3}}}\right)
\end{eqnarray}
and reduce \eqref{4} to the following equation:
\begin{eqnarray}
\ddot{\alpha}=-{\alpha}({\omega_0^{2}}+h\cos{2\omega
t})-q\dot{\alpha} \label{5}
\end{eqnarray}
which is just a Mathieu equation if one sets damping to zero.

The presence of parametric resonance in \eqref{5} is examined in
Ref. \cite{Landau} for driving frequencies $\omega$ close to
pendulum oscillation frequency $\omega_0$. Actually, similar
analysis could be done for arbitrary $\omega$ and the existence of
parametric resonance in the system will cause undamped oscillations,
chaos and some more interesting phenomena. In order to find out what
conditions should be fulfilled for this to occur, we should seek the
solution of equation \eqref{5} in the following form:
\begin{eqnarray}
{\alpha}=a(t)\cos{\omega t}+b(t)\sin{\omega t}
\end{eqnarray}
considering $a(t)$ and $b(t)$ as slow functions of time and
neglecting their second derivatives, \eqref{5} is simplified to the
following form:
\begin{eqnarray}
X\cos{\omega t}+Y\sin{\omega t}=0
\end{eqnarray}
Where coefficients $X$ and $Y$ both depend on $a(t)$ and $b(t)$. For
the equation to be true, both coefficients should be equal to zero.
Thus we get a set of two equations, where our goal is to find the
regimes of parametric resonance. For this, we should seek for the
solution in the exponential form
$a(t)=Ae^{st}$ and $b(t)=Be^{st}$ and two equations are derived:
\begin{eqnarray}
A\cdot(2s\omega+q\omega)-B\cdot({\omega_0}^2+\frac{h}{2}-{\omega}^2)=0
\nonumber \\
A\cdot({\omega}^2-\frac{h}{2}-{\omega_0}^2)-B\cdot(2s\omega+q\omega)=0.
\label{6}
\end{eqnarray}
Finally we get from the compatibility condition:
\begin{eqnarray}
s=\frac{{\omega_0}^2+\frac{h}{2}-{\omega}^2-q\omega}{2\omega}
\label{7}
\end{eqnarray}
Considering parametric instability growth rate $s$ to be positive,
the instability condition will be:
\begin{eqnarray}
h\geq 2\mid w^{2}-{w_0}^{2}+2q\omega\mid. \label{66}
\end{eqnarray}

This defines the limits of existence of parametric resonance and its
dependence on various parameters, but all of these is valid for
small angles. In order to get the full dynamics we should solve
differential equation \eqref{3} in a full range of angles. $F_x$ and $F_y$ components of magnetic force are known
from derivative of dipole-dipole energy. If we do not consider the
angle as small, we will not be able to make the approximations that
has been done before. In general, $F_x$ and $F_y$ components are very
complicated expressions and it is impossible to solve the equation
\eqref{3} analytically. Therefore I performed numerical simulations
using Matlab.

\noindent{\it Numerical simulations:} Our next goal is to prove
theoretically the existence of chaos in the system, considering
deviation angles as arbitrary. The given equation of motion \eqref{3}
has been solved using $ode45$ toolbox of Matlab program with an
initial guess that chaos should occur when parametric resonance for
small angles takes place. And this appears to be true, because as
the numerical simulations show, when there is parametric instability
in the system, it is always chaotic. To prove the existence of
chaos, the common way is to check, whether changing any parameter
insignificantly, the difference between the first and second
measurement of some variable increases exponentially in time. In
other words, Lyapunov exponent should be calculated in order to
analyze the behavior of chaotic motion. To calculate the exponent,
one has to deviate e.g. initial angle $\alpha(0)$ by small value
making it $\alpha^\prime(0)$ and as time evolves, divide the
resulting difference between angles $\alpha(t)$ and
$\alpha^\prime(t)$ on initial deviation. Taking out logarithm from
this, dividing on time and averaging the results upon the initial
deviations Lyapunov exponent of the process could be defined.
Positive exponent is an obvious indication of the presence of chaos and one
should look at the simultaneous presence of parametric resonance
condition in the system.

\begin{figure}[t]
\includegraphics[scale=.65]{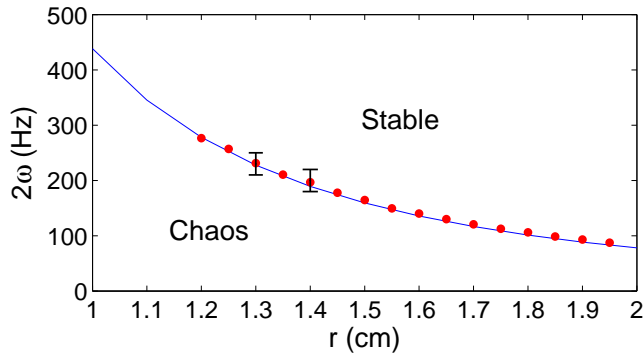}
\caption{Comparison of theoretical and numerical results for finding
the boundaries of parametric resonance and chaos. Solid line is
theoretical curve according to \eqref{66} and red dots are plotted
using numerical simulations. Error bars show the boundary area of
chaos in experiments.} \label{fig_02}
\end{figure}

Another test to check the relation between parametric resonance and
chaos in our case of magnetic pendulum is to look whether the system
performs large angle oscillations starting from initial
insignificant deviations. In other words, if we give the pendulum
very small initial angle, for example 0.0001 rad, and after
some time it starts to oscillate with normal angles, this means that
there is parametric resonance and chaos in the system. The latter
scenario is observed in experiments when the system is in
parametrically unstable regime. In Fig. \ref{fig_02} solid blue line
indicates theoretical boundary line of parametric resonance, so it
is also boundary of chaos and stability. Red dots are boundaries of
chaos from numerical simulations, so the discrepancy between theory
and numerical calculations is really small. We also indicate by
error bar experimental range where transition from stability to
chaos occurs.

\begin{figure}[t]
\includegraphics[scale=.57]{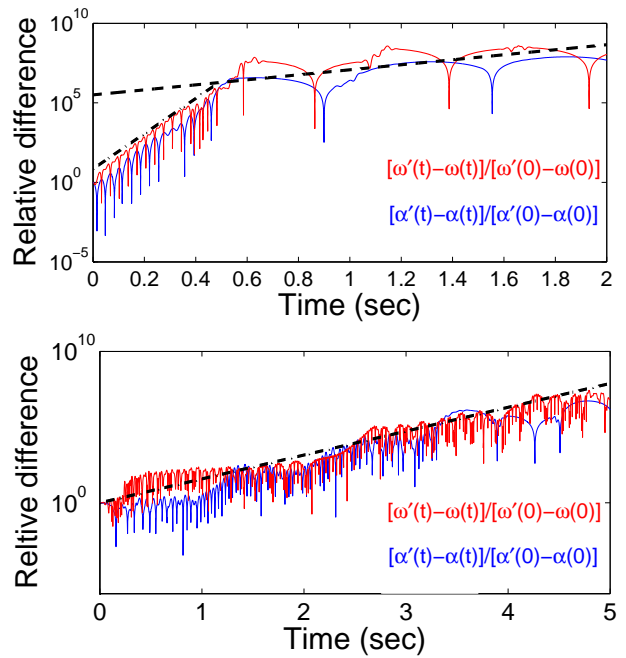}
\caption{In both graphs blue and red solid lines show relative angle
and angular velocity differences, respectively, versus time. Black
dashed line follows from theoretical estimate of parametric
resonance growth rate \eqref{7} equal to $s=3.63$. Driving frequency
in both graphs is $29$ Hz and pendulum parameters are indicated in
the text. Initial angle is very small $0.0001$ rad. In case of upper
panel initial large growth rate with the exponent $\approx 23$
(dotted-dashed line) is caused by the fact that at $t=0$ current
starts to flow in solenoid instantly. Bottom graph shows the same
relative differences of angles and velocities, but when we multiply
dipolar moment of solenoid on the factor $1-\exp(-t)$, it
prevents current (and therefore force) to gain large values almost
instantly.} \label{fig_03}
\end{figure}

\begin{figure}[t]
\includegraphics[scale=.55]{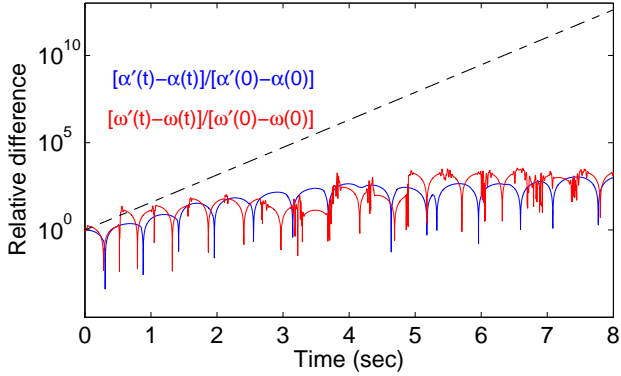}
\caption{The same as in Fig. \ref{fig_03} except initial angle,
which is 1 rad. This is not a small angle and that is why numerical
results do not fit with theoretical growth rate indicated by dashed
line. Lyapunov exponent in this case is $1.55$ and theoretical
growth rate is $s=3.63$.} \label{fig_04}
\end{figure}

In Figs. \ref{fig_03} and \ref{fig_04} the evolution of relative
differences of the pendulum angle $\alpha$ and angular velocity
$\dot\alpha$ ($\omega$) are presented. For instance, in case of
relative angle difference we use for its calculation the formula
$\left[\alpha^\prime(t)-\alpha(t)\right]/\left[\alpha^\prime(0)-\alpha(0)\right]$.
We evaluate the dynamics from two small initial values, e.g.
$\alpha(0)=0.0001$ and $\alpha^\prime(0)=0.0001001$ and
average upon different initial deviations. Lyapunov exponent
values are as follows: for angles we get the exponent value equal to $3.47$ and for
angular velocities it is $3.56$ which in good approximation
coincides with parametric instability growth rate $s=3.63$,
calculated from formula \eqref{7}. As seen from upper graph of Fig.
\ref{fig_03}, in the beginning we have rapid growth of relative
angle and velocity differences, which is characterized by a value of
Lyapunov exponent equal to $23$, and it is quite different than
theoretical growth rate. This effect happens because in numerical
experiments at $t=0$ current starts to flow in solenoid abruptly,
therefore a force of finite value instantly appears on magnet, and
this is the cause of strange behaviour of pendulum. In order to exclude
such a scenario we multiply the dipolar moment of solenoid
${L_1}(t)\equiv {L^0_1}\cos{2\omega t}$ on the time dependent factor
$1-\exp(-t)$, modelling smooth growth of the current in the solenoid.
One can observe the result on the bottom panel of Fig. \ref{fig_03}.
As seen, no rapid growth takes place in the beginning of the time,
because, the current (and therefore force) starts to increase slowly
and the value of the Lyapunov exponent coincides with theoretical
growth rate (black dashed line).

No calculations of Lyapunov exponent were made using scenario
displayed on the bottom panel, it is only used to prove and explain
the reason of rapid growth in the upper panel of Fig. \ref{fig_03}.
In calculations of Lyapunov exponent of angles and velocities the
beginning of time where rapid growth takes place has not been
considered, and calculated Lyapunov exponent coincides with
theoretical growth rate $s=3.63$. Theoretical and numerical results
are really close in Fig. \ref{fig_03}, that is because of the fact
that the initial deviation angle of the pendulum is small.

While in Fig. \ref{fig_04} the initial deviation angle is around $1$
rad and Lyapunov exponent is equal to $1.55$ and it does not
coincide with theoretical growth rate $s=3.63$. This fact was
predictable, because pendulum actually spends little time at small
deviation angles where parametrical instability is in force.
However, the range of stabilty-chaos diagram plotted in Fig.
\ref{fig_02} is still valid for large initial deviation angles.

\begin{figure}[b]
\includegraphics[scale=.27]{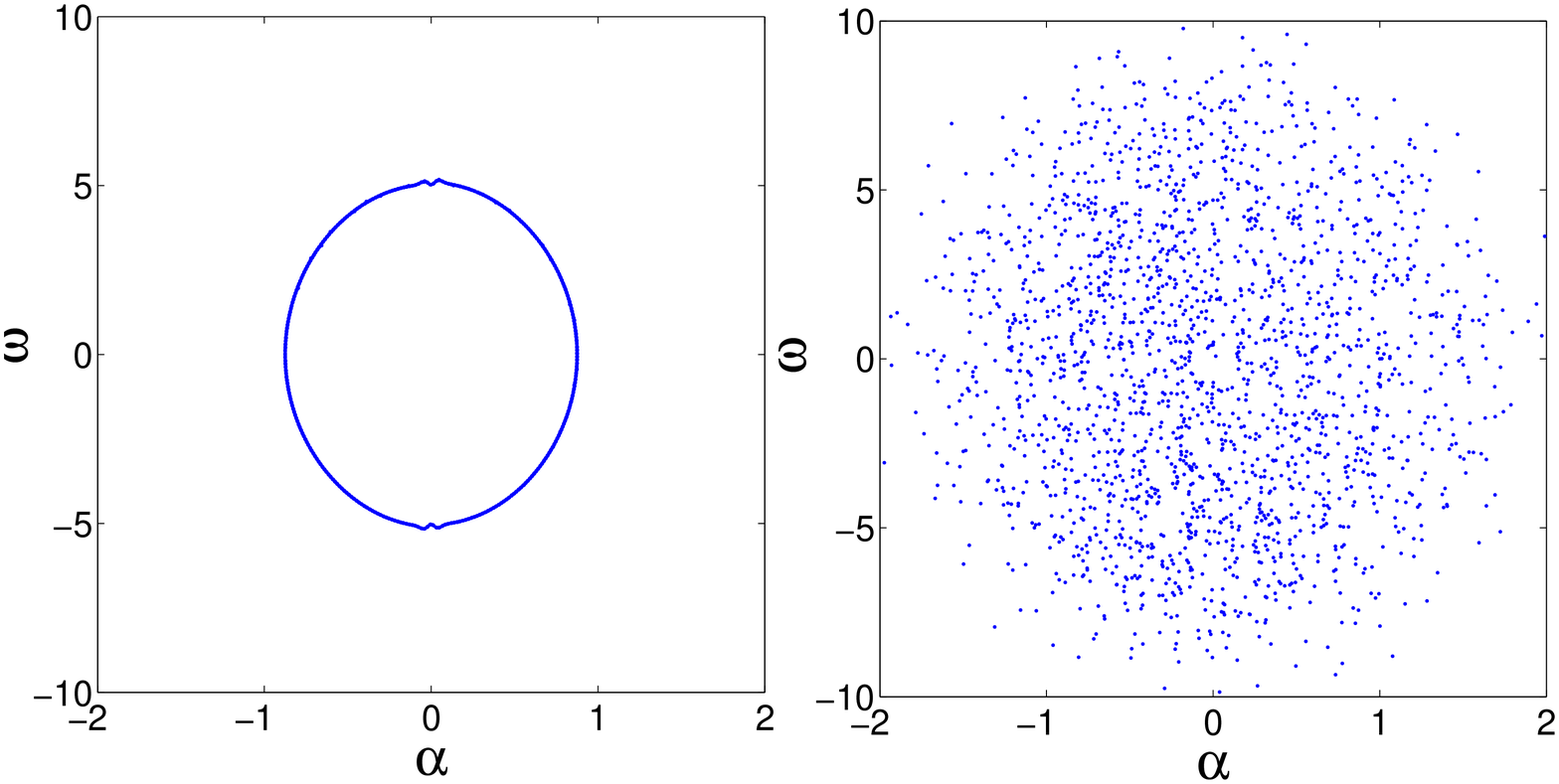}
\caption{Left: Poincare graph when the system is not chaotic
corresponding to the minimal distance between solenoid and magnet
$r_0=22$ mm. Right: Poincare graph in case of chaos when
$r_0=14$ mm. The initial angle in both cases is large
($\alpha(0)=1$ rad), and time step is the period of oscillations of
free pendulum with an initial $1$ rad angle. Other parameters of the
system are given in the text.} \label{fig_05}
\end{figure}

Here are the parameters for figures \ref{fig_03} and \ref{fig_04}:
minimal distance between magnet and solenoid is $r_0=14 mm$, the
length of the pendulum is taken as $\ell=28$ cm, dipolar moment
amplitude of solenoid is $L_1^0=1.8$ $A\cdot{m^2}$, dipolar moment
of the magnet is $L_2=0.2$ $A\cdot{m^2}$, mass of the magnet is
$m=0.05$ kg, damping coefficient is taken as $q=0.01$ and ac current
frequency is $2\omega= 58\pi$.

In Fig. \ref{fig_05} two Poincare graphs are displayed, both of them
express the system dynamics with the same initial parameters, except
minimal distance from solenoid to the magnet: for left graph
$r_0=22$ mm, which corresponds to the regular evolution, while at
the right the chaotic behavior is observed for $r_0=14$ mm. In
both cases the initial angle is large (1 rad) and the time period of
plotting dots is the own oscillation period of the free pendulum
with the same parameters. If the time period of plotting dots were the oscillation period in the presence of magnetic field and not the own oscillation period of the free pendulum, Poincare graph would be a point for the parameters of left graph of Fig 5 (non-chaotic regime). Poincare graphs are only displayed for the cases where the initial angle is large, because in case of small
angles it is clear that when parametric resonance occurs, the system
is unavoidably chaotic.

\begin{figure}[t]
\includegraphics[scale=.57]{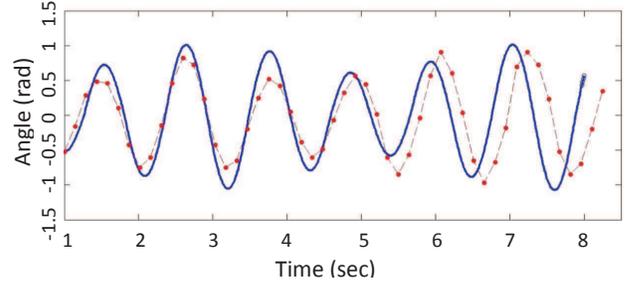}
\caption{Comparison of theoretical model (blue solid line) and
experimental one in case of regular dynamics. Red dots are
experimental data and red dashed line is for guide of eyes. }
\label{fig_06}
\end{figure}

\noindent{\it Experiments:} In Fig. \ref{fig_01} experimental
magnetic pendulum is displayed. All parameters are easily measurable
except dipolar moments of solenoid and magnet. For this purpose we
have used magnetic field sensor and the measurements were made in
different locations (more than $50$ locations). Applying then
regression formula we have estimated values of dipolar moments. The
experimental parameters which are used in numerical simulations are
given in the previous section. While conducting experiments slow
motion camera has been used in order to track pendulum motion.
After the data has been processed on the computer and points have
been plotted on theoretical graph (see Fig. \ref{fig_06}). It has
been taken into account that experimental pendulum is not a
mathematical one, and thus \eqref{3} has been rewritten for the
physical pendulum case. Besides that, experimentally, damping
proportional to the velocity is to be taken into account.
The damping coefficient $q$ was measured as follows: the time needed for damping
from initial angle is recorded, and then it is compared to numerical
calculations made in Matlab with different damping coefficients. For
the coefficient $q=0.01$ the damping took the same time as in the
experiment. In Fig. \ref{fig_06} a comparison of theoretical model
and experiment has been made in case of nonchaotic regime and one
can clearly see that theory and experiment is well-fitted. The
difference between them of course grows with time, because there are
some experimental errors, which constantly act on the motion
characteristics. The main error is that dipoles in reality have
size, especially solenoid while in theoretical model we have made an
assumption that they are point-like.

The video in supplemental material is recorded for the case when the
system is chaotic. We have zero initial deviation. When we let the
current flow into the solenoid the pendulum start large amplitude
oscillations. From a very small initial angle system stars large
amplitude oscillations, so it proves the existence of parametric
resonance and consequently the chaos in the system.

\noindent{\it Conclusions:} We have proved and examined the
existence and interrelation of parametric resonance and chaos in the
system of magnetic pendulum. Lyapunov exponents were calculated
using numerical simulations and were compared with theoretical
growth rate. Lyapunov exponent for small angles (angular velocities)
matches with theoretical growth rate and for large angles it is
different, as it was expected. The overall conclusion is that our
magnetic pendulum system is chaotic only when the conditions for
parametric resonance are fulfilled. Besides that, experiments have
been carried out and give a very good agreement with theoretical model and numerical
simulations.

I would like to give special thanks to T. Gachechiladze and G.
Mikaberidze for very useful discussions.

\end{document}